\documentstyle[psfig]{lamuphys}
\makeatletter
\let\chapter\hid@chapter
\makeatother
\begin{document}
\pagenumbering{arabic}
\title{Dust and Molecules at High Redshift} 

\author{F. Combes}

\institute{Observatoire de Paris, DEMIRM, \\
61 Av. de l'Observatoire, F-75 014 Paris, France}

\maketitle

\begin{abstract}
In the last years, progress has been very rapid in the domain of molecules
at high redshift, and we know in better detail now the molecular content
in several systems beyond $z=1$ and up to $z \sim 5$, either through
millimeter and sub-millimeter emission lines, continuum dust emission,
or millimeter absorption lines in front of quasars.
 The first discovery  in 1992 by Brown and van den Bout of CO lines
at $z=2.28$ in a gravitationally lensed starburst galaxy, strongly
stimulated searches of other systems, but these were harder than foreseen,
and less than 10 other systems have been discovered in CO emission.
Redshifts range between 2 and 5, the largest being BR1202-0725
at $z=4.69$. Most of these systems, if not all, are gravitationally
amplified objects. Some have been discovered first through their
dust emission, relatively easy to detect because of the
negative K-correction effect. The detection of all these systems
could give an answer about the debated question of the star-formation rate
as a function of redshift. The maximum of star-formation rate, found
around $z=2$ from optical studies, could shift to higher $z$ if the most
remote objects are hidden by dust.

Absorption in front of quasars can also probe cold gas at high redshift, 
taking advantage of very high spatial
($\le$ 10$^{-3}$ arcsec) and spectral (30m/s) resolutions,
and sampling column-densities
between N(H$_2$)= 10$^{20}$ et 10$^{24}$ cm$^{-2}$.
Up to now four absorption systems have been discovered in the millimeter range,
and up to 20 molecular lines have been detected in a single object
(HCO$^+$, HNC, HCN, N$_2$H$^+$, C$^{18}$O, CS, H$_2$CO, CN, etc....).
From the diffuse components, one can measure the 
cosmic black body temperature as a function of redshift. The high
column densities component allow to observe important molecules
not observable from the ground, like O$_2$, H$_2$O and LiH for example.

All these preliminary studies will be carried out at large scales
with future millimeter instruments, and some perspectives are given.
\end{abstract}
\section{Detection of CO emission lines at high redshift}

The first detection of millimeter CO lines in emission in the Faint IRAS source
F10214+4724 at $z = 2.28$  by Brown \& Vanden Bout (1991, 1992) 
created a surprise,
since it was a redshift 30 times larger than that of the most
distant CO emission discovered in a galaxy.  The H$_2$ mass derived was
reaching 10$^{13} h^{-2}$ M$_\odot$, with the standard CO-H$_2$ conversion
ratio, a huge mass although the FIR to CO luminosities was still
compatible with that of other more nearby starbursts. Since then, the 
derived H$_2$ mass has been reduced by large factors, both with better
data and realizing that the source is amplified through a gravitational 
lens by a large factor (Solomon et al 1992, 1997).

After the first discovery, many searches for other candidates took place,
but they were harder than expected, and only a few,
often gravitationally amplified,
objects have been detected: the lensed Cloverleaf quasar
H 1413+117 at $z=2.558$ (Barvainis et al. 1994),
the lensed radiogalaxy MG0414+0534 at $z=2.639$ (Barvainis et al. 1998),
the possibly magnified object
BR1202-0725 at $z=4.69$ (Ohta et al. 1996, Omont et al. 1996),
the amplified submillimeter-selected hyperluminous galaxy SMM02399-0136
(Frayer et al. 1998), at $z=2.808$, and the magnified BAL quasar APM08279+5255,
at $z=3.911$, where the gas temperature derived from the CO lines is
$\sim$ 200K, maybe excited by the quasar (Downes et al. 1998).
Recently Scoville et al. (1997) reported the detection of the first
non-lensed object at $z=2.394$, the weak radio galaxy 53W002,
and Guilloteau et al. (1997) the radio-quiet quasar BRI 1335-0417, at $z=4.407$,
which has no direct indication of lensing.
If the non-amplification is confirmed, these objects
 would contain the largest molecular contents known
(8-10 10$^{10}$ M$_\odot$ with a standard CO/H$_2$
conversion ratio, and even more
if the metallicity is low).
The derived molecular masses are so high that H$_2$ would constitute
between 30 to 80\% of the total dynamical mass (according to the unknown
inclination), if the standard CO/H$_2$ conversion ratio was adopted.
The application of this conversion ratio is however doubtful, and it is
possible that the involved H$_2$ masses are 3-4 times lower (Solomon
et al. 1997). Results are summarized in Table 1.

\begin{table}[h]
\caption[ ]{CO data for high redshift objects}
\begin{flushleft}
\begin{tabular}{lllclcl}  
\hline
Source    & $z$   &  CO  & S  & $\Delta$V& MH$_2$   & Ref  \\
          &       &line  & mJy  & km/s & 10$^{10}$ M$_\odot$    &           \\
\hline
F10214+4724 & 2.285 & 3-2  & 18 & 230  & 2$^*$      &  1   \\
53W002      & 2.394 & 3-2  &  3 & 540  & 7      &  2   \\
H 1413+117  & 2.558 & 3-2  & 23 & 330  & 6      &  3   \\
MG 0414+0534& 2.639 & 3-2  &  4 & 580  & 5$^*$      &  4   \\
SMM 02399-0136&2.808& 3-2  &  4 & 710  & 8$^*$      &  5   \\
APM 08279+5255&3.911& 4-3  &  6 & 400  & 0.3$^*$    &  6   \\
BR 1335-0414& 4.407 & 5-4  &  7 & 420  & 10         &  7   \\
BR 1202-0725& 4.690 & 5-4  &  8 & 320  & 10         &  8   \\
\hline
\end{tabular}
\end{flushleft}
$^*$ corrected for magnification, when estimated\\
Masses have been rescaled to $H_0$ = 75km/s/Mpc. When multiple images
are resolved, the flux corresponds to their sum\\
(1) Solomon et al. (1992), Downes et al (1995); (2) Scoville et al. (1997);
(3) Barvainis et al (1994); (4) Barvainis et al. (1998); (5) Frayer et al.
(1998); (6) Downes et al. (1998); (7) Guilloteau et al. (1997);
(8) Omont et al. (1996)
\end{table}

Surprisingly, very few starburst galaxies have been detected in
the CO lines at intermediate redshifts (between 0.3 and 2), although many have
been observed (e.g Yun \& Scoville 1998, Lo et al 1999). A possible explanation
is the lower probability of magnification by lenses in this range. Figure 1 is
summarizing these results.

\begin{figure}
\psfig{figure=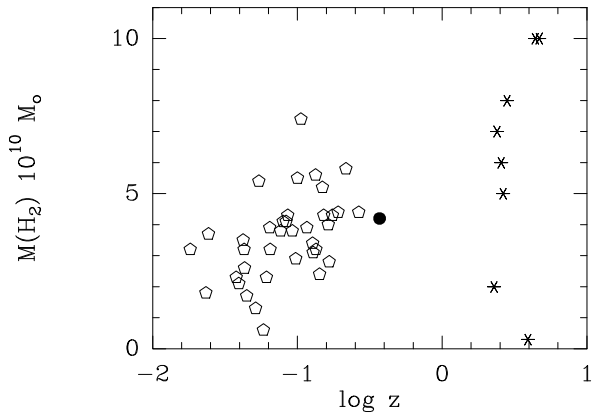,bbllx=2cm,bblly=1cm,bburx=12cm,bbury=14cm,width=8cm,angle=-90}
\caption{Derived H$_2$ masses for the CO-detected objects at high redshift (filled
stars), compared to the ultra-luminous-IR  sample of Solomon et al (1997, open pentagons).
There is no detected object between 0.3 and 2.2 in redshift, except the quasar 
3c48, marked as a filled dot (Scoville et al 1993, Wink et al 1997). The dispersion of points
at high $z$ is due to uncertainties in the actual magnification, that allows
low masses to be detected. }
\end{figure}

\section{Detection of dust emission}

Most of the previous sources, detected in the CO lines, had previously been detected
in the dust continuum. At high redshift, it becomes easier to detect the dust emission,
because of the large K-correction (e.g. Blain \& Longair 1993): the emission is roughly
varying as $\nu^4$ with the frequency $\nu$ in the millimeter range,
until the maximum around 60$\mu$m. At one mm, it is even easier to detect objects
at $z = 5$ than $z = 1$. This has motivated deep searches in blank fields with
sensitive instruments, since they should be dominated by high redshift objects,
if they exist in sufficient numbers. 

The first search was made with the SCUBA bolometer
on JCMT (Hawaii) towards  a cluster of galaxies, thought to serve as a gravitational lens
for high-$z$ galaxies behind (Smail et al 1997). The amplification is in average a factor 2.
A large number of sources were found, all at large redshifts ($z > 1$), extrapolated 
to 2000 sources per square degree (above 4mJy), 
revealing a large positive evolution with redshift,
i.e. an increase of starbursting galaxies. Searches toward the Hubble Deep Field-North
(Hughes et al 1998), and towards the Lockman hole and SSA13 (Barger et al 1998),
have also found a few sources, allowing to derive a similar density of sources:
800 per square degree, above 3 mJy at 850 $\mu$m. This already can account for 
50\% of the cosmic infra-red background (CIRB), that has been estimated by
Puget et al (1996) and Hauser et al (1998) from COBE data.
The photometric redshifts of these sources range between 1 and 3.
Their identification with optical objects might be  uncertain (Richards 1998).
However, Hughes et al (1998) claim that the star formation rate derived from
the far-infrared might be in some cases 10 times higher than derived
from the optical, due to the high extinction. If  only some of the sources
have a redshift higher than 4, it will flatten the Madau curve at high $z$.

Eales et al (1999) surveyed some of the CFRS fields at 850$\mu$m
 with SCUBA and found also that the sources can account for a significant
fraction of the CIRB background ($\sim$ 30\%). Their interpretation
in terms of the star formation history is however slightly different, in that they
do not exclude that the submm luminosity density could evolve in the same
way as the UV one.
Deep galaxy surveys at 7 and 15$\mu$m with ISOCAM also see an evolution
with redshift of star-forming galaxies: heavily extincted starbursts represent less
than 1\% of all galaxies, but 18\% of the star formation rate out to $z = 1$ 
(Flores et al 1999).

\section{Molecules in absorption}

Molecules can also be detected in absorption in the millimeter range,
and a few systems have been discovered at high redshift, between
$z$ = 0.2 to 1 in the last years (Wiklind \& Combes 1994, 95, 96;
Combes \& Wiklind 1996). The 
sensitivity is such that a molecular cloud on the line of sight
of only a few solar masses is enough to detect a signal,
while in emission, upper limits at the same distance are of the order 
of 10$^{10}$ M$_\odot$. Some general properties of the known
absorbing systems are summarised in Table 2.
They reveal to be the continuation
at high column densities (10$^{21}$--10$^{24}$ cm$^{-2}$)
of the whole spectrum of absorption systems, from the
Ly$\alpha$ forest (10$^{12}$--10$^{19}$ cm$^{-2}$)
to the damped Ly$\alpha$ and HI 21cm absorptions
(10$^{19}$--10$^{21}$ cm$^{-2}$).

\begin{table}[h]
\caption{Properties of molecular absorption line systems at high $z$} 
\begin{center} 
\begin{tabular}{lcccccrc}
\hline
Source & z$_{\rm a}^{a}$ & z$_{\rm e}^{b}$ &
$N_{\rm CO}$ & $N_{\rm H_2}$ & $N_{\rm HI}$ &
A$_{V}^{\prime c}$ & $N_{\rm HI}/N_{H_2}$ \\
 & & & cm$^{-2}$ & cm$^{-2}$ & cm$^{-2}$ & & \\
\hline \\
PKS1413+357   & 0.24671 & 0.247  & $2.3 \times 10^{16}$ & $4.6 \times 10^{20}$
& $1.3 \times 10^{21}$ & 2.0 & 2.8 \\
B3\,1504+377A & 0.67335 & 0.673  & $6.0 \times 10^{16}$ & $1.2 \times 10^{21}$
& $2.4 \times 10^{21}$ & 5.0 & 2.0 \\
B3\,1504+377B & 0.67150 & 0.673   & $2.6 \times 10^{16}$ & $5.2 \times 10^{20}$
& $<7 \times 10^{20}$ & $<$2 & $<$1.4 \\
B\,0218+357   & 0.68466 & 0.94   & $2.0 \times 10^{19}$ & $4.0 \times 10^{23}$
& $4.0 \times 10^{20}$ & 850 & $1 \times 10^{-3}$ \\
PKS1830--211A & 0.88582 & 2.51      & $2.0 \times 10^{18}$ & $4.0 \times 10^{22}$
& $5.0 \times 10^{20}$ & 100 & $1 \times 10^{-2}$ \\
PKS1830--211B & 0.88489 & 2.51      & $1.0 \times 10^{16 d}$ &
$2.0 \times 10^{20}$ & $1.0 \times 10^{21}$ & 1.8 & 5.0 \\
PKS1830--211C & 0.19267 & 2.51      & $<6 \times 10^{15}$                   &
$<1 \times 10^{20}$ & $2.5 \times 10^{20}$ & $<$0.2 & $>$2.5 \\
\hline
\end{tabular}
\end{center}
$^{a}${Redshift of absorption line}
$^{b}${Redshift of background source}
$^{c}${Extinction corrected for redshift using a Galactic extinction law}
$^{d}${Estimated from the HCO$^{+}$ column density
of $1.3 \times 10^{13}$\,cm$^{-2}$}
$^{e}${21cm HI data taken from Carilli et al. 1992,
1993, 1998}
\end{table}

About 15 different molecules have been detected
in absorption at high redshifts, in a total of 30
different transitions. This allows a detailed
chemical study and comparison with local clouds
(Wiklind \& Combes 1997).
Up to now, no significant variations in abundances
have been found as a function of redshift, at
least within the large intrinsic dispersion
already existing within a given galaxy.
Note that the high redshift allows us to
detect some new molecular lines, never observed from the ground
at $z = 0$, such as O$_2$ (Combes et al 1997), H$_2$O or LiH (Combes
\& Wiklind 1997, 1998).

Among the four detected systems, there are two cases of
confirmed gravitational lenses, with multiple images
(the two other cases are likely to be internal absorption).
This is not unexpected, since detection of molecular absorption
requires that the line of sight to the background QSO must
pass close to the center of an intervening galaxy. The
surface density is then larger than the critical one for
multiple images. In the present two cases (B0218+357 and
PKS1830-210), the separation between the two images is small
enough that only a galaxy bulge is required for the lensing.
This makes these two objects good candidates for the
determination of the Hubble constant (Wiklind \& Combes
1995, 1998).
The narrowness of absorption lines and their high redshift make
them good candidates to refine the constraints on the variation of
coupling constants with cosmic time, by comparing
HI and molecular lines for example (Drinkwater et al 1998
and ref therein). However, the various lines  may
come with different velocities, with different weights
(see fig. 2).

\begin{figure}
\psfig{figure=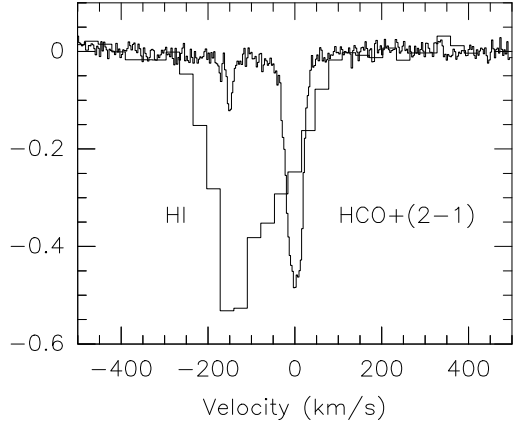,bbllx=2cm,bblly=1cm,bburx=12cm,bbury=14cm,width=8cm,angle=-90}
\caption{ Absorption lines observed towards PKS1830-21, at a redshift of
$z = 0.88582$. HCO$^+$(2-1) and HI (lower V-resolution) are compared, 
in arbitrary units. The HI
component (Chengalur et al 1999) is much stronger towards the V = -150km/s
component than all the detected molecular lines.}
\end{figure}

Molecular clouds are usually very cold, with a kinetic temperature
of the order of 10-20K. The excitation temperature of the
molecules could be even colder, close to the background temperature
$T_{bg}$. This is the case when the absorption occurs in diffuse gas,
where the density is not enough to excite the rotational ladder of
the molecules.
This is precisely the case of the gas absorbed in front of PKS1830-211, where
$T_{ex} \sim T_{bg}$ for most of the molecules. The measurement of
$T_{ex}$ requires the detection of two nearby transitions. When the
lower ones is optically thick, only an upper limit can be derived
for $T_{ex}$. Ideally, the two transitions should be optically thin,
but  then the higher one is very weak, and long integration times are
required.

The results obtained with the SEST-15m and IRAM-30m on different 
molecules agree and 
are plotted as a single point in
figure 3. Surprisingly, the bulk of measurements
points towards an excitation temperature lower than the background
temperature at $z=0.88582$, i.e. $T_{bg}$= 5.20K. 
We suspect that this could be due to a variation of the total 
continuum flux, due to a micro-lensing event.

\begin{figure}
\psfig{figure=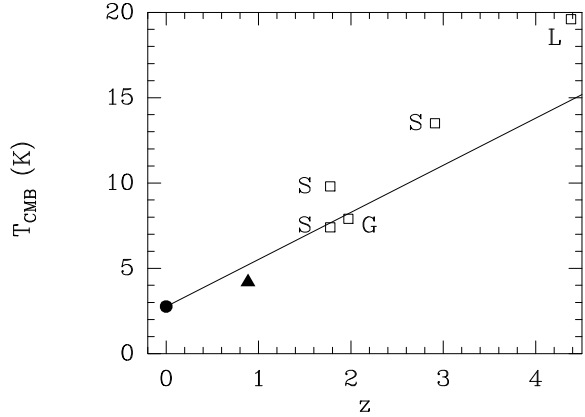,bbllx=2cm,bblly=1cm,bburx=12cm,bbury=14cm,width=8cm,angle=-90}
\caption{Summary of CMB temperature measurements as a function
of redshift. The filled dot is from COBE (Mather et al 1994). The squares are
upper limits obtained on the CI or CII from Songaila et al (1994ab, S), Lu et al (1996, L)
and Ge et al (1997, G). Our point is the filled triangle. The line is the (1+z) expected
variation. }
\end{figure}

\section{Perspectives with future instruments}

Knowing the principal characteristics of detected starbursts, it will be easy to detect
them at much higher redshifts with the planned future millimeter instruments, even
without the help of gravitational lensing (as today).  As described in section 1, 
molecular masses range from 10$^{10}$ to 10$^{11}$ M$_\odot$, the dust temperature
is high, about 30-50K (up to 100K), and their size is strikingly small,
below one kpc (300pc disks, Solomon et al 1997). In these conditions, the average
column density of H$_2$ is  10$^{24}$ cm$^{-2}$, and the dust becomes optically
thick at $\lambda <$ 150$\mu$m.  Two extreme simple models can be made
about the corresponding molecular medium: either the gas is distributed in an
homogeneous sphere, at a temperature of 50K, with a density of  10$^{3}$ cm$^{-3}$
in average, or the gas is clumpy, distributed in cold dense clouds, with embedded
hot cores. The expected flux coming from such starbursting regions 
in the various CO lines, and the continuum flux from the dust emission, can then
be derived as a function of redshift (cf Combes et al 1999). An example is given
in figure 4.

\begin{figure}[h]
\psfig{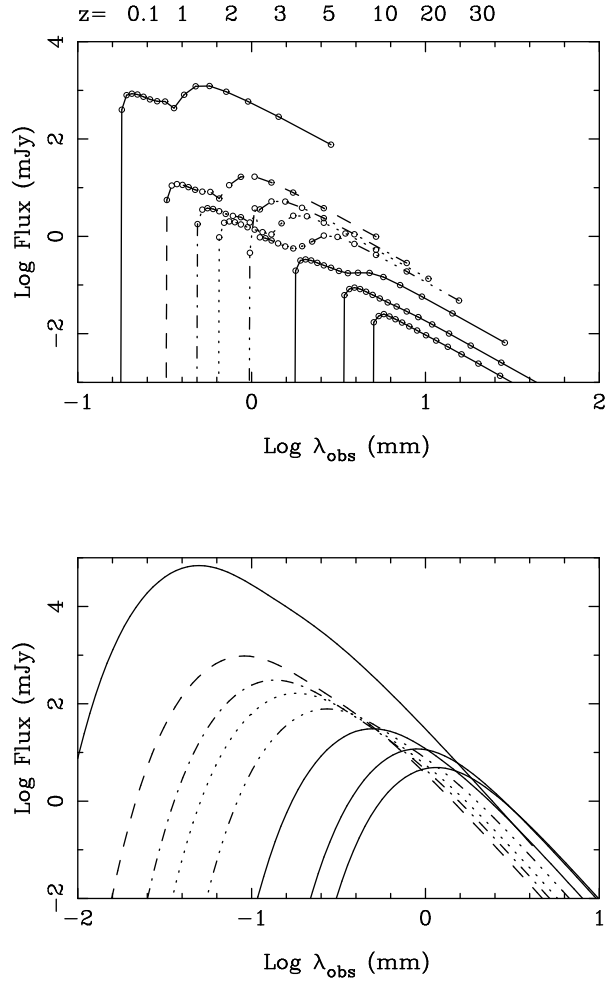}
\caption{ Expected flux for a starburst model in two-component clouds, for various
redshifts $z$ = 0.1, 1, 2, 3, 5, 10, 20, 30, and $q_0$ = 0.5.
Top are the CO lines, materialised
each by a circle. Bottom is the continuum emission from dust.}
\end{figure}

It can be seen in this picture that the CO lines are less favoured at high redshift
than the continuum emission; the K-correction is less because the CO lines are
optically thick. It is already much more difficult to detect objects at $z = 5$
than at $z = 1$, contrary to the dust emission.
The most favourable wavelengths to detect the CO lines are always
longer than 1mm (assuming that the kinetic temperature of the gas is the
same as the dust temperature).
While it is already possible to detect in the continuum such starbursts at any
redshifts (up to 30), it will be possible with the upcoming future millimeter
instruments (the Green-Bank-100m of NRAO, the LMT-50m of UMass-INAOE,
the LSA-MMA (Europe/USA) and the LMSA (Japan) interferometers) to detect
them in the CO lines, up to the highest redshifts, without gravitational
magnification.

%

\end{document}